\documentclass[a4paper,12pt]{article}

\pdfoutput=1

\usepackage[tbtags]{amsmath}
\usepackage{amssymb}
\usepackage{ifthen}
\usepackage{slashed}
\usepackage{amsfonts}
\usepackage{mathrsfs}
\usepackage{bm}
\usepackage{graphicx,subfigure,booktabs}
\usepackage[numbers,sort&compress]{natbib}
\usepackage{verbatim}
\usepackage{appendix}

\usepackage[colorlinks,
            linkcolor=black,
            filecolor=black,
            anchorcolor=black,
            urlcolor=black,
            citecolor=blue
            ]{hyperref}

\usepackage{color}
\usepackage{ulem}
\usepackage{setspace}
\numberwithin{equation}{section}

\newlength{\dinwidth}
\newlength{\dinmargin}
\makeatletter
\newcommand{\thickhline}{%
    \noalign {\ifnum 0=`}\fi \hrule height 1pt
    \futurelet \reserved@a \@xhline
}
\makeatother

\allowdisplaybreaks

\begin{document}

\title{\bf Revisiting the radiative decays \boldmath{$J/\psi \rightarrow \gamma\eta^{(\prime)}$} in perturbative QCD}

\author{Jun-Kang He\footnote{hejk@mails.ccnu.edu.cn}\; and
Ya-Dong Yang\footnote{yangyd@mail.ccnu.edu.cn}\\[15pt]
\small Institute of Particle Physics and Key Laboratory of Quark and Lepton Physics~(MOE), \\
\small Central China Normal University, Wuhan, Hubei 430079, China}
\date{}



\maketitle
\vspace{0.2cm}

\begin{abstract}
{\noindent}In the framework of perturbative QCD, the radiative decays $J/\psi\rightarrow\gamma\eta^{(\prime)}$ are revisited
in detail, where the involved one-loop integrals are evaluated analytically with the light quark masses kept. We have found that
the sum of loop integrals is insensitive to the light quark masses and the branching ratios $\mathcal{B}(J/\psi\rightarrow\gamma\eta^{(\prime)})$ barely depend on the shapes of $\eta^{(\prime)}$ distribution amplitudes. With the parameters of $\eta-\eta^{\prime}$ mixing extracted from low energy processes and $J/\psi\rightarrow\gamma\eta^{(\prime)}$ by means of nonperturbative matrix elements $\langle0|G_{\mu\nu}^a\tilde{G}^{a,\mu\nu}|\eta^{(\prime)}\rangle$ based on $U_{A}(1)$ anomaly dominance argument, we could not give the ratio $R_{J/\psi}$ in agreement with experimental result. However, using the parameters, especially the mixing angle $\phi=33.5^{\circ}\pm0.9^{\circ}$, extracted from $\gamma^{\ast}\gamma-\eta^{\prime}$ transition form factor measured at $q^{2}=112~\mathrm{GeV}^{2}$ by BaBar collaboration, we obtain $R_{J/\psi}=4.70$ in good agreement with $R_{J/\psi}^{exp}=4.65\pm0.21$. As a crossing check, with $\Gamma^{exp}(\eta^{(\prime)}\rightarrow\gamma\gamma)$ and our results for $J/\psi\rightarrow\gamma\eta^{(\prime)}$, we get $\phi=33.9^{\circ}\pm0.6^{\circ}$. The difference between the determinations of $\phi$ is briefly discussed.
\end{abstract}

\newpage

\section{Introduction}
\label{sec:intro}

Since its discovery\cite{Aubert:1974js,Augustin:1974xw}, the $J/\psi$ meson has always been an active topic in particle physics~\cite{Brambilla:2004wf,Brambilla:2010cs,Voloshin:2007dx}. The heavy quarkonium physics might be crucially important to
improve our understanding of quantum chromodynamics (QCD), especially, the interplay of perturbative quantum chromodynamics (pQCD) with nonperturbative QCD. The Okubo-Zweig-Iizuka (OZI)-forbidden radiative decays of $J/\psi\rightarrow\gamma\eta^{(\prime)}$ are expected to proceed predominantly via two virtual gluons which subsequently convert to $\eta^{(\prime)}$, with the photon emitted from the initial charm quarks. Such decays have been the subject of many experimental and theoretical studies~\cite{Brambilla:2004wf,Brambilla:2010cs,Voloshin:2007dx}. Theoretically, these decays provide a clean environment to study the conversion of gluons into hadrons. In this respect, the radiative decays $J/\psi\rightarrow\gamma\eta^{(\prime)}$ are of particular interest, since they are also closely related to the issues of $\eta-\eta^{\prime}$ mixing, which are important ingredients for understanding many interesting phenomena related to the $\eta$ and $\eta^{\prime}$ mesons.

In the literature, the exclusive radiative decays $J/\psi\rightarrow\gamma\eta^{(\prime)}$ have been studied in different approaches. The decay widths $\Gamma(J/\psi\rightarrow\gamma\eta^{(\prime)})$ were calculated by Novikov {\it et al.}~\cite{Novikov:1979uy}, with the assumption that these decays occur as a consequence of the $U_A(1)$ anomaly and are, therefore, controlled by the nonperturbative gluonic matrix elements $\langle0|G_{\mu\nu}^a\tilde{G}^{a,\mu\nu}|\eta^{(\prime)}\rangle$, where $G_{\mu\nu}^a$ is the gluon field strength tensor and
$\tilde{G}^{a,\mu\nu}=\frac{1}{2}\epsilon^{\mu\nu\alpha\beta}G_{\alpha\beta}^a$ its dual tensor. Then the ratio of the decay widths takes the form~\cite{Novikov:1979uy}
\begin{eqnarray}\label{key formula}
R_{J/\psi}=\frac{\Gamma(J/\psi\rightarrow\gamma\eta^{\prime})}{\Gamma(J/\psi\rightarrow\gamma\eta)}
=\left(\frac{M_{J/\psi}^{2}-m_{\eta^{\prime}}^{2}}{M_{J/\psi}^{2}-m_{\eta}^{2}}\right)^{3}
{\left|\frac{\langle0|G_{\mu\nu}^a\tilde{G}^{a,\mu\nu}|\eta^{\prime}\rangle}{\langle0|G_{\mu\nu}^a\tilde{G}^{a,\mu\nu}|\eta\rangle}\right|}^{2},
\end{eqnarray}
where the matrix elements $\langle0|G_{\mu\nu}^a\tilde{G}^{a,\mu\nu}|\eta^{(\prime)}\rangle$ can be calculated with the QCD sum rules and other approaches. For example, Chao~\cite{Chao:1989pi,Chao:1990im} and Kuang {\it et al.}~\cite{Kuang:1990kd} derived the expressions of the matrix elements $\langle0|G_{\mu\nu}^a\tilde{G}^{a,\mu\nu}|\eta^{(\prime)}\rangle$
in the large-$N_{c}$ approach and QCD multipole expansion, respectively.

In the Feldmann-Kroll-Stech (FKS) scheme for $\eta-\eta^{\prime}$ mixing system~\cite{Feldmann:1998vh}, the nonperturbative matrix elements $\langle0|G_{\mu\nu}^a\tilde{G}^{a,\mu\nu}|\eta^{(\prime)}\rangle$ can also be expressed as function of the phenomenological parameters $f_{q}$, $f_{s}$, $\phi$, and the ratio $R_{J/\psi}$ reads
\begin{eqnarray}
R_{J/\psi}=\tan^{2}\phi\frac{m_{\eta^\prime}^{4}}{m_{\eta}^{4}}\left(\frac{M_{J/\psi}^{2}-m_{\eta^{\prime}}^{2}}
{M_{J/\psi}^{2}-m_{\eta}^{2}}\right)^{3},
\end{eqnarray}
where $\phi$ denotes the mixing angle of $\eta-\eta^{\prime}$ system. Compared $R_{J/\psi}$ with its experimental
value, the mixing angle is found to be $\phi=39.3^{\circ}\pm1.0^{\circ}$~\cite{Feldmann:1998vh}. However, as it had been pointed out that the above equation was calculated in the approximation with the assumption of ground state dominance and neglection of continuum contributions to the dispersion relations~\cite{Novikov:1979uy,Ball:1995zv}. It is also noticed that the matrix elements $\langle0|G_{\mu\nu}^a\tilde{G}^{a,\mu\nu}|\eta^{(\prime)}\rangle$ induced by the $U_{A}(1)$ anomaly are a higher twist effect(twist-4)~\cite{Chernyak:1983ej}. The $\langle0|G_{\mu\nu}^a\tilde{G}^{a,\mu\nu}|\eta^{(\prime)}\rangle$ would give the main contributions to the radiative decays $J/\psi\rightarrow\gamma\eta^{(\prime)}$, only in the case of the leading twist contributions were strongly suppressed. Although the leading twist contributions from the gluonic content of $\eta^{(\prime)}$ are suppressed by a factor of $m_{\eta^{(\prime)}}^{2}/M_{J/\psi}^{2}$~\cite{Baier:1981pm}, the assumption that the matrix elements $\langle0|G_{\mu\nu}^a\tilde{G}^{a,\mu\nu}|\eta^{(\prime)}\rangle$ dominate the radiative decays $J/\psi\rightarrow\gamma\eta^{(\prime)}$ could be broken, because the leading twist contributions from the quark-antiquark content of $\eta^{(\prime)}$ are not suppressed much at the energy scale of $M_{J/\psi}$.

The first pQCD investigation of these decays were carried out by K\"{o}rner {\it et al.}~\cite{Korner:1982vg} about thirty years ago. They took the annihilation of $c\bar{c}$ quarks to be a short-distance process described by pQCD, and nonperturbative dynamics of the bound states factorized to wave functions. It has been argued that pQCD asymptotic behaviors may be expected at the scale of $M_{J/\psi}$~\cite{Duncan:1980qd,Brodsky:1981kj,Chernyak:1981zz,Chernyak:1983ej}. However, in this pioneer work~\cite{Korner:1982vg}, the nonrelativistic quark model with the weak-binding approximation, has been taken for both $J/\psi$ and $\eta^{(\prime)}$. Whereafter, the nonrelativistic approximation for $\eta^{(\prime)}$ wave functions in~\cite{Korner:1982vg}
was improved by K\"{u}hn~\cite{Kuhn:1983yr} with light-cone expansion. However, in the calculation of the loop integrals, the approximation of $m_{\eta^{(\prime)}}^{2}/M_{J/\psi}^{2}\approx0$ was made~\cite{Kuhn:1983yr}. In our calculation, we would keep $m_{\eta^{(\prime)}}$ and $m_{u,d,s}$, which result in our final analytical expression of the loop function is much complicated than the ones in Refs.~\cite{Korner:1982vg,Kuhn:1983yr}. We notice that our results can reproduce the one in Ref.~\cite{Kuhn:1983yr} in the limit of $m_{\eta^{(\prime)}}^{2}/M_{J/\psi}^{2}\rightarrow0$, and detail comparison is presented in the Appendix. Recently, several groups have revisited $J/\psi\rightarrow\gamma\eta^{(\prime)}$ in the framework of pQCD~\cite{Ma:2002ww,Yang:2004wy,Li:2005ug,Gao:2006,Li:2007dq}. In works~\cite{Yang:2004wy,Li:2005ug}, the light-cone distribution amplitudes (DAs) were adopted for $\eta^{(\prime)}$. However, in Ref.~\cite{Li:2005ug}, the decay widths of $J/\psi\rightarrow\gamma\eta^{(\prime)}$ were found to be very sensitive to the light quark masses of $\eta^{(\prime)}$ involved in the loop integrals, which is needed to be made clear, since the sensitivities of the loop integrals to light quark masses usually point to the possible infra-red (IR) divergences.

Besides the aforementioned QCD approaches, the radiative decays $J/\psi\rightarrow\gamma\eta^{(\prime)}$ have also been studied with phenomenological models, such as the approach with an effective lagrangian~\cite{Gerard:2013gya} and the approach considering the $\eta_{c}-\eta^{(\prime)}$ mixings~\cite{Chao:1989pi,Chao:1990im,Zhao:2010mm}. Generally, predictions compatible with the experimental measurements could be obtained.

In this paper, we present a detail calculation of these decays in the framework of pQCD. The bound-state property of $J/\psi$ is parameterized by its Bethe-Salpeter (B-S) wave function, while the $\eta$ and $\eta^{\prime}$ are described by their light-cone DAs. Then the loop integrals are evaluated analytically with the light quark masses kept.

In our calculation, the loop function is found to be insensitive to the light quark masses, which differs from the results in
Ref.~\cite{Li:2005ug}. Moreover, our results of the $\mathcal{B}(J/\psi\rightarrow\gamma\eta^{(\prime)})$ are also insensitive to the shapes of the light meson DAs. The theoretical uncertainties due to choices
of different $\eta^{(\prime)}$ DAs available in the literature are negligible in the prediction for the ratio $R_{J/\psi}$, so that, the mixing angle of $\eta-\eta^{\prime}$ mixing could be reliably extracted.
In addition, the corrections from QED processes $J/\psi\rightarrow\gamma^{\ast}\rightarrow\gamma\eta^{(\prime)}$ are also considered in our calculation.

The paper is organized as follows. In section~\ref{sec:framework}, we present the formalism for calculating the decay amplitudes of $J/\psi\rightarrow\gamma\eta^{(\prime)}$. Numerical results are presented in section~\ref{sec:numerical analysis}, and the final part is our summary. The analytical expressions for the dimensionless key function $H_{0}$ and discussions of its properties are presented in the Appendix.

\section{The radiative decays $J/\psi\rightarrow\gamma\eta^{(\prime)}$ in pQCD}
\label{sec:framework}

\subsection{The contributions of the quark-antiquark content of $\eta^{(\prime)}$}
\label{subsec:QCDq}

\begin{figure}[!!htb]
\centering
\includegraphics[width=0.45\textwidth]{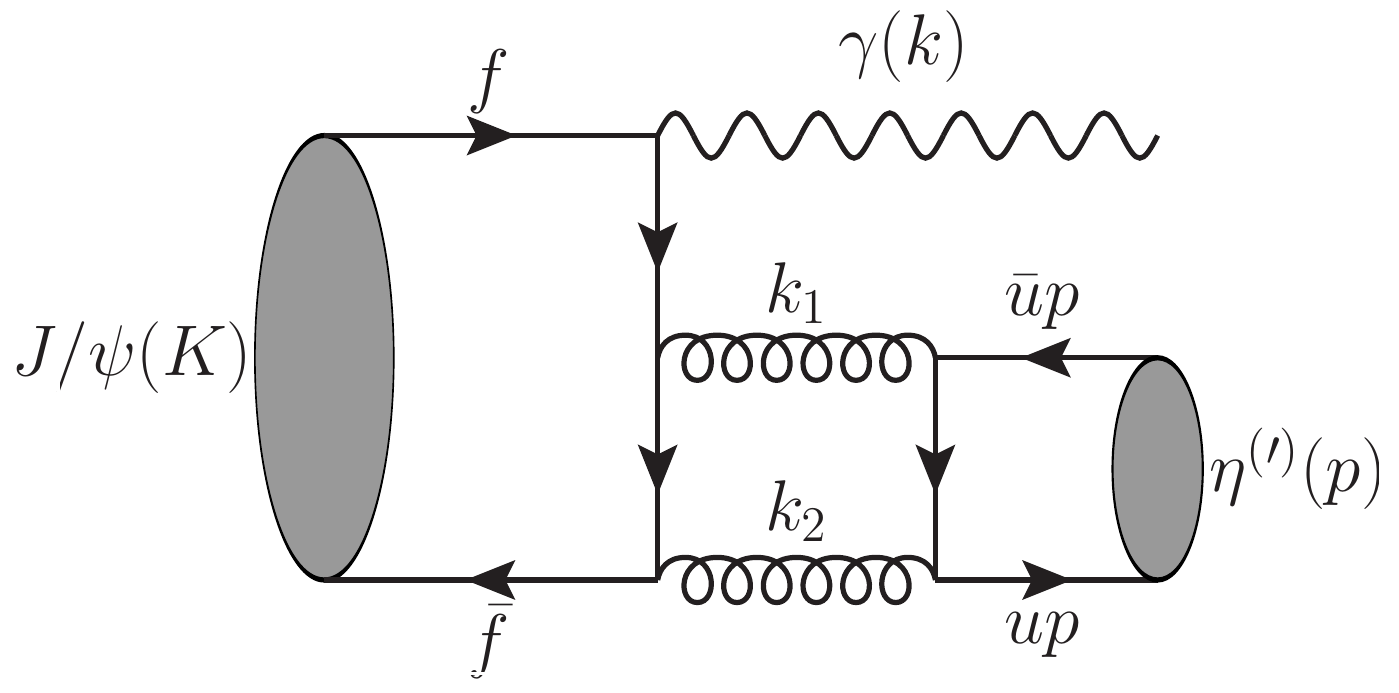}
\caption{\label{jpsirp}One typical Feynman diagram for $J/\psi\rightarrow\gamma\eta^{(\prime)}$ with the quark-antiquark content of $\eta^{(\prime)}$. Kinematical variables are labelled.}
\end{figure}

For the quark-antiquark content of $\eta^{(\prime)}$, the leading order contributions to the radiative decays $J/\psi\rightarrow\gamma\eta^{(\prime)}$ arise from one-loop QCD processes. The corresponding Feynman diagram is illustrated in Fig.~\ref{jpsirp}. There are other five Feynman diagrams from permutations of the photon and gluon legs. Usually, it is convenient to divide the invariant amplitude into two parts. One part describes the
effective coupling between $J/\psi$, a real photon and two virtual gluons, the other part describes the effective coupling between pseudoscalar meson $\eta^{(\prime)}$ and two virtual gluons.
To evaluate these effective couplings, we need to deal with the nonperturbative effects of the mesons. Generally, factorization is employed. For the heavy $J/\psi$, we still use the weak-binding approximation, and factorize the nonperturbative bound-state effects into its B-S wave function. For the light mesons $\eta$ and $\eta^{\prime}$, we use their light-cone DAs.

In the rest frame of $J/\psi$, the amplitude of $J/\psi\rightarrow \gamma g^{\ast}g^{\ast}$ can
be decomposed into hard-scattering part and the B-S wave function of $J/\psi$~\cite{Resag:1993xq}
\begin{eqnarray}\label{A}
A&=&A^{\alpha\beta\mu\nu}\varepsilon_{\alpha}(K)\epsilon_{\beta}^{\ast}(k)\epsilon_{\mu}^{\ast}(k_{1})\epsilon_{\nu}^{\ast}(k_{2})\nonumber\\
&=&-i\sqrt{3}\int\frac{\mathrm{d}^{4}q_{c}}{(2\pi)^{4}}\mathrm{Tr}\left[\chi(q_{c})\hat{\mathcal{O}}(q_{c})\right],
\end{eqnarray}
where $\chi(q_{c})$ is the B-S wave function of $J/\psi$ and $\sqrt{3}$ is the color factor. $K$ and $\varepsilon(K)$ stand for the momentum and the polarization vector of the $J/\psi$, $k$, $k_{1}$, $k_{2}$ and $\epsilon(k)$, $\epsilon(k_{1})$, $\epsilon(k_{2})$ stand for momenta and polarization vectors of the photon and the gluons, respectively. The momenta of the $c$ and $\bar{c}$ quarks are parameterized as
\begin{eqnarray}
f=\frac{K}{2}+q_{c},~~~~~~~   \bar{f}=\frac{K}{2}-q_{c},
\end{eqnarray}
where $q_{c}$ is the relative momentum between the $c$ and $\bar{c}$ quarks.
Since the B-S wave function $\chi(q_{c})$ is sharply peaked when $q_{c}\sim 0$ for the heavy $J/\psi$ in the nonrelativistic limit, one may neglect the $q_{c}$-dependence in the hard-scattering amplitude $\hat{\mathcal{O}}(q_{c})$ in the leading order approximation, then the amplitude can be simplified to
\begin{eqnarray}
A&=&-i\sqrt{3}\int\frac{\mathrm{d}^{4}q_{c}}{(2\pi)^{4}}\mathrm{Tr}\left[\chi(q_{c})\hat{\mathcal{O}}(0)\right]\nonumber\\
&=&-\sqrt{3}\int\frac{\mathrm{d}^{3}\boldsymbol{q_{c}}}{(2\pi)^{3}}\mathrm{Tr}\left[
\psi(\boldsymbol{q_{c}})\hat{\mathcal{O}}(0)\right],
\end{eqnarray}
where the Salpeter function is defined as
\begin{eqnarray}
\psi(\boldsymbol {q_{c}})=\frac{i}{2\pi}\int\mathrm{d}q_{c}^{0}\chi(q_{c}).
\end{eqnarray}
For the vector meson $J/\psi$, the Salpeter function with leading order Dirac structures reads\cite{Wang:2005qx,Negash:2015rua}
\begin{eqnarray}\label{salpeter}
\psi(\boldsymbol{q_{c}})=\phi(\boldsymbol{q_{c}}^{2})(M+\slashed{K})\slashed{\varepsilon}(K),
\end{eqnarray}
where $\phi(\boldsymbol{q_{c}}^{2})$ is a scalar function of $\boldsymbol{q_{c}}^{2}$, and $M$ represents the mass of $J/\psi$. With the help
of the definition of the $S$-wave wave function evaluated at the origin~\cite{Guberina:1980dc}
\begin{eqnarray}\label{zeroapp}
\int\frac{\mathrm{d}^{3}\boldsymbol{q_{c}}}{(2\pi)^{3}}\phi(\boldsymbol{q_{c}}^{2})=\frac{1}{2}\sqrt{\frac{1}{4\pi M}}R_{\psi}(0),
\end{eqnarray}
one can obtain~\cite{Korner:1982vg}
\begin{eqnarray}\label{A}
A^{\alpha\beta\mu\nu}\varepsilon_{\alpha}(K)\epsilon_{\beta}^{\ast}(k)\epsilon_{\mu}^{\ast}(k_{1})\epsilon_{\nu}^{\ast}(k_{2})=-\frac{1}{2}\sqrt{\frac{3}{4\pi M}}R_{\psi}(0)
\mathrm{Tr}\left[(M+\slashed{K})\slashed{\varepsilon}(K)\hat{\mathcal{O}}(0)\right].
\end{eqnarray}
And the hard-scattering amplitude $\hat{\mathcal{O}}(0)$ can be written as
\begin{eqnarray}
\hat{ \mathcal{O}}(0)&=&iQ_{c}eg_{s}^{2}\frac{\delta_{ab}}{6}\slashed{\epsilon}^{\ast}(k_{2})
 \frac{\slashed{k}_{2}-\slashed{k}-\slashed{k}_{1}+M}{-2 (k+k_{1})\cdot k_{2}}\slashed{\epsilon}^{\ast}(k)\frac{\slashed{k}_{2}+\slashed{k}-\slashed{k}_{1}+M}{-2 (k+k_{2})\cdot k_{1}}
      \slashed{\epsilon}^{\ast}(k_{1})\nonumber\\
      & &+\textrm{(5~permutations~of~ $k_{1}$,~$k_{2}$~and~$k$)},
\end{eqnarray}
where we have made the nonrelativistic approximation $M\approx 2m_{c}$.

\begin{figure}[!!htb]
\centering
\includegraphics[width=0.45\textwidth]{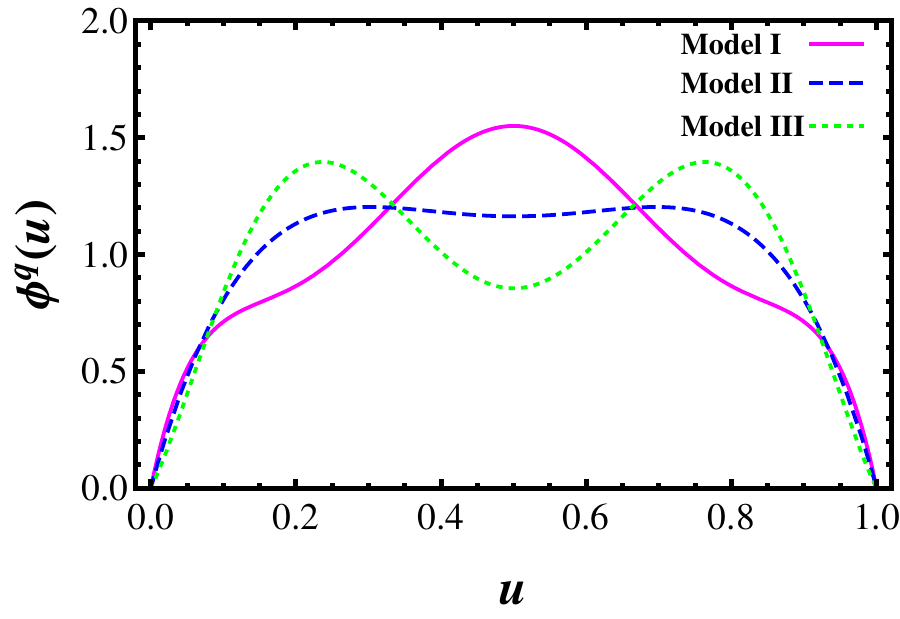}
\caption{\label{DAs}The shapes of the corresponding DAs at the scale of $\mu=m_{c}$ }
\end{figure}

The light-cone expansion of the matrix elements of the meson $\eta^{(\prime)}$ over quark and antiquark fields reads~\cite{Chernyak:1983ej}
\begin{eqnarray}
\langle \eta^{(\prime)}(p) |\bar{q}_{\alpha}(x)q_{\beta}(y)|0\rangle &=&\frac{i}{4}f_{\eta^{(\prime)}}^{q}\left(\slashed{p}\gamma_{5}\right)_{\beta\alpha}
 \int\mathrm{d}u e^{i(\bar{u}p\cdot y+up\cdot x)}\phi^{q}(u)+\cdots,
\end{eqnarray}
where the high twist terms are omitted. With this definition, one can obtain the coupling of $g^{\ast}g^{\ast}-\eta^{(\prime)}$~\cite{Muta:1999tc,Yang:2000ce,Ali:2000ci}:
\begin{eqnarray}
M^{\mu\nu}&=&-i (4\pi \alpha_{s})\delta_{ab}\epsilon^{\mu\nu\rho\sigma}k_{1\rho}k_{2\sigma}\nonumber\\
& &\times\sum_{q=u,d,s}\frac{f_{\eta^{(\prime)}}^{q}}{6}\int^{1}_{0}du\phi^{q}(u)
\left(\frac{1}{\bar{u}k_{1}^{2}+uk_{2}^{2}-u\bar{u}m^{2}-m_{q}^{2}}+(u\leftrightarrow\bar{u})\right),
\end{eqnarray}
where $u$ is the momentum fraction carried by the quark and $\bar{u}=1-u$, $m_{q}$ is the mass of the quark($q=u,d,s$), $m$ is the mass of $\eta^{(\prime)}$.  The light-cone DA is~\cite{Agaev:2014wna}
\begin{eqnarray}
\phi^{q}(u)&=&\phi_{AS}(u)\left(1+\sum_{n=2,4\cdots}c^{q}_{n}(\mu)C_{n}^{\frac{3}{2}}(2u-1)\right)
\end{eqnarray}
with the asymptotic form of DA $\phi_{AS}(u)=6u(1-u)$ and $c^{q}_{n}(\mu)$ the Gegenbauer moments.
In Table~\ref{tab:coefficients}, we list three models of the DAs discussed in Ref.~\cite{Agaev:2014wna}. Their shapes are shown in Fig.~\ref{DAs}, where $c_{n}^{q}$ are evaluated at $\mu=m_{c}$.

\begin{table}[!!htb]
  \caption{\label{tab:coefficients}Gegenbauer coefficients of three sample models at the scale of $\mu_{0}=1~\mathrm{GeV}$.}
\vspace{0.2cm}
\centering
  \begin{tabular}{l c r @{.} l c}
  \hline\hline
   Model    ~~~&~~~$c_{2}^{q}(\mu_{0})$~~&\multicolumn{2}{c}{~~~~$c_{4}^{q}(\mu_{0})$}~~~&~~~$c_{2}^{g}(\mu_{0})$    \\
  \hline
  I         ~~~&~~~$0.10$~~~&~~~$0$&$10$ ~~~&~~~$-0.26$ \\
  II        ~~~&~~~$0.20$~~~&~~~$0$&$00$ ~~~&~~~$-0.31$ \\
  III       ~~~&~~~$0.25$~~~&~~~$-0$&$10$~~~&~~~$-0.25$ \\
  \hline\hline
  \end{tabular}
\end{table}

The decay amplitude of $J/\psi\rightarrow \gamma\eta^{(\prime)}$ can be obtained by multiplying the above two couplings, inserting the gluon propagators and performing the loop integrations
\begin{equation}
     M_{T}=T^{\alpha\beta}\varepsilon_{\alpha}(K)\epsilon^{\ast}_{\beta}(k)=\frac{1}{2}\int\frac{\mathrm{d}^4k_{1}}{(2 \pi)^4}A^{\alpha\beta\mu\nu}M_{\mu\nu}\varepsilon_{\alpha}(K)\epsilon^{\ast}_{\beta}(k)\frac{i}{k_{1}^2+i\epsilon}
     \frac{i}{k_{2}^2+i\epsilon},
\end{equation}
where the factor $1/2$ takes into account that the two gluons have already been interchanged both in $A^{\alpha\beta\mu\nu}$ and $M_{\mu\nu}$.
Using parity conservation, Lorentz invariance and gauge invariance, it can be proved that
\begin{eqnarray}\label{Tmn}
T^{\alpha\beta}\sim \epsilon^{\alpha\beta\mu\nu}p_{\mu}k_{\nu},
\end{eqnarray}
so there is only one independent helicity amplitude $H_{QCD}^{q}$~\cite{Korner:1982vg}
\begin{eqnarray}\label{Tee}
T^{\alpha\beta}\varepsilon_{\alpha}(K)\epsilon^{\ast}_{\beta}(k)&=&H_{QCD}^{q}h^{\alpha\beta}\varepsilon_{\alpha}(K)\epsilon^{\ast}_{\beta}(k),
\end{eqnarray}
where
\begin{eqnarray}
h^{\alpha\beta}&=&\frac{i}{p\cdot k}\epsilon^{\alpha\beta\mu\nu}p_{\mu}k_{\nu}.
\end{eqnarray}
With the help of the helicity projector~\cite{Korner:1982vg}
\begin{eqnarray}
\mathbb{P}^{\alpha\beta}&=&\frac{1}{2}h_{\alpha^{\prime}\beta^{\prime}}
\left(-g^{\alpha\alpha^{\prime}}+\frac{K^{\alpha}K^{\alpha^{\prime}}}{M^{2}}\right)\left(-g^{\beta\beta^{\prime}}\right)\nonumber\\
&=&-\frac{i}{2p\cdot k}\epsilon^{\alpha\beta\mu\nu}p_{\mu}k_{\nu},
\end{eqnarray}
one can obtain the helicity amplitude
\begin{eqnarray}\label{scalar}
H_{QCD}^{q}&=&T^{\alpha\beta}\mathbb{P}_{\alpha\beta}\nonumber\\
&=&\frac{2Q_{c}}{9}\sqrt{4\pi\alpha_{e}}(4\pi\alpha_{s})^{2}\sqrt{\frac{3}{\pi M}}R_{\psi}(0)\sum_{q=u,d,s}\frac{f_{\eta^{(\prime)}}^{q}}{M}H_{q}.
\end{eqnarray}
The dimensionless function $H_{q}$ reads
\begin{eqnarray}\label{scahq}
H_{q}&=&-\frac{1}{16\pi^{2}}\frac{2}{1-x}\int\mathrm{d}u\phi^{q}(u)I(u)
\end{eqnarray}
with $x=m^{2}/M^{2}$. $I(u)$ is the sum of the loop integrals of the six Feynman diagrams for the decays
\begin{eqnarray}
I(u)&=&\frac{1}{i \pi^{2}}\int\frac{\mathrm{d}^{4}q}{16}
\Bigg{(}\frac{N_{1}}{C_{1}D_{2}D_{3} D_{4}D_{5}}+\frac{N_{2}}{C_{1}D_{1} D_{3} D_{4} D_{5}}+\frac{N_{3}}{D_{1}D_{2} D_{3} D_{4} D_{5}}\nonumber\\
& &+\frac{N_{4}}{C_{1}D_{1}D_{3} D_{4} D_{5}}+\frac{N_{5}}{C_{1}D_{2}D_{3}D_{4} D_{5}}+\frac{N_{6}}{D_{1}D_{2}D_{3}D_{4} D_{5}}\Bigg{)}+(u\leftrightarrow\bar{u})
\end{eqnarray}
with $q=k_{1}-k_{2}$. Here the expressions of the denominators are given by
\begin{eqnarray}
C_{1}&=&\frac{1}{4}\left[(p-k)^{2}-M^{2}\right]+i\epsilon,\nonumber\\
D_{1}&=&\frac{1}{4}\left[(q-k)^{2}-M^{2}\right]+i\epsilon,\nonumber\\
D_{2}&=&\frac{1}{4}\left[(q+k)^{2}-M^{2}\right]+i\epsilon, \nonumber\\
D_{3}&=&\frac{1}{4}\left[(q+(\bar{u}-u)p)^{2}-4m_{q}^{2}\right]+i\epsilon,\nonumber\\
D_{4}&=&\frac{1}{4}(q+p)^{2}+i\epsilon,\nonumber\\
D_{5}&=&\frac{1}{4}(q-p)^{2}+i\epsilon,
 \end{eqnarray}
and the six numerators $N_{i}$ read
\begin{eqnarray}
N_{1}&=&N_{5} = \frac{1}{4} k\cdot p \left[k\cdot q \left(m^2-p\cdot q\right)+k\cdot p \left(q^2-p\cdot q\right)\right],\nonumber\\
N_{2}&=&N_{4}= \frac{1}{4} k\cdot p \left[k\cdot p \left(p\cdot q+q^2\right)-k\cdot q \left(m^2+p\cdot q\right)\right],\nonumber\\
N_{3}&=&N_{6}= \frac{1}{4} \left[2 k\cdot p k\cdot q p\cdot q-m^2 k\cdot q^2-q^2 k\cdot p^2\right].
\end{eqnarray}
The third and the sixth terms in $I(u)$ are five-point loop integrals, and the other terms are four-point loop integrals
since the denominator $C_{1}$ is independent of the loop momentum $q$. Before going to calculate the integral, we would present a short analysis of its IR properties. When one of the gluons goes on-shell, i.e., $q\rightarrow p$, the denominators $D_{2}$, $D_{3}$(with $u=1-m_{q}/m$), $D_{5}$ tend to zero. Following to the procedure in Ref.~\cite{Dittmaier:2003bc}, we can find that the one-loop integrals in the individual Feynman diagram for $J/\psi\rightarrow\gamma\eta^{(\prime)}$ have soft singularities, which depend on the momentum fraction $u$, and can make the convolution integral over $u$ become sensitive to the shapes of the $\eta^{(\prime)}$ DAs. Moreover, the divergent term due to light quark pole would result in the final numerical results strongly dependent on the light quark masses. However, summing up the six Feynman diagrams, one can obtain
\begin{eqnarray}\label{Iu}
I(u)&=&\frac{1}{i \pi^{2}}\int\frac{\mathrm{d}^{4}q}{16}\frac{(m^{2}-q^{2})\left(k\cdot p q^{2}-k\cdot q p\cdot q\right)}{4 D_{1}\,D_{2}\,D_{3}\,D_{4}\,D_{5}}+(u\leftrightarrow\bar{u}).
\end{eqnarray}
When $q= p+\lambda$ ($\lambda$ is a small quantity), the numerator, which arises from the summation of all Feynman diagrams,
\begin{eqnarray}
(m^{2}-q^{2})\left(k\cdot p q^{2}-k\cdot q p\cdot q\right)\sim \lambda^{2},
\end{eqnarray}
and the on-shell propagators
\begin{eqnarray}
D_{2}\simeq\frac{1}{2}K\cdot \lambda\sim \lambda,~~~~~~~ D_{3}\simeq\frac{m_{q}}{m}p\cdot\lambda\sim \lambda,~~~~~~~ D_{5}\simeq\frac{1}{4}\lambda^{2}\sim \lambda^{2}.
\end{eqnarray}
For the ultrasoft gluon($\lambda\rightarrow 0$), the contributions to the loop integral have the form
\begin{eqnarray}
\int\limits_{q= p+\lambda}\mathrm{d}^{4}q\frac{(m^{2}-q^{2})\left(k\cdot p q^{2}-k\cdot q p\cdot q\right)}{ D_{1}\,D_{2}\,D_{3}\,D_{4}\,D_{5}}&\sim&\int\mathrm{d}^{4}\lambda\frac{\lambda^{2}}{\lambda^{4}}
\rightarrow 0.
\end{eqnarray}
It means the sum of the loop integrals is IR safe.

\begin{figure}[!!htb]
\centering
\includegraphics[width=0.47\textwidth]{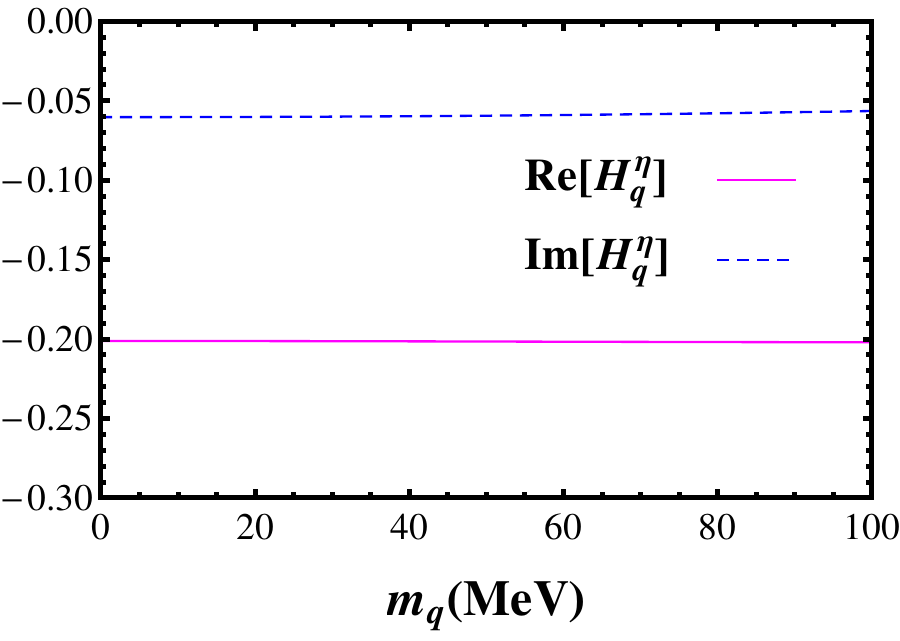}\hspace{0.5cm}
\includegraphics[width=0.47\textwidth]{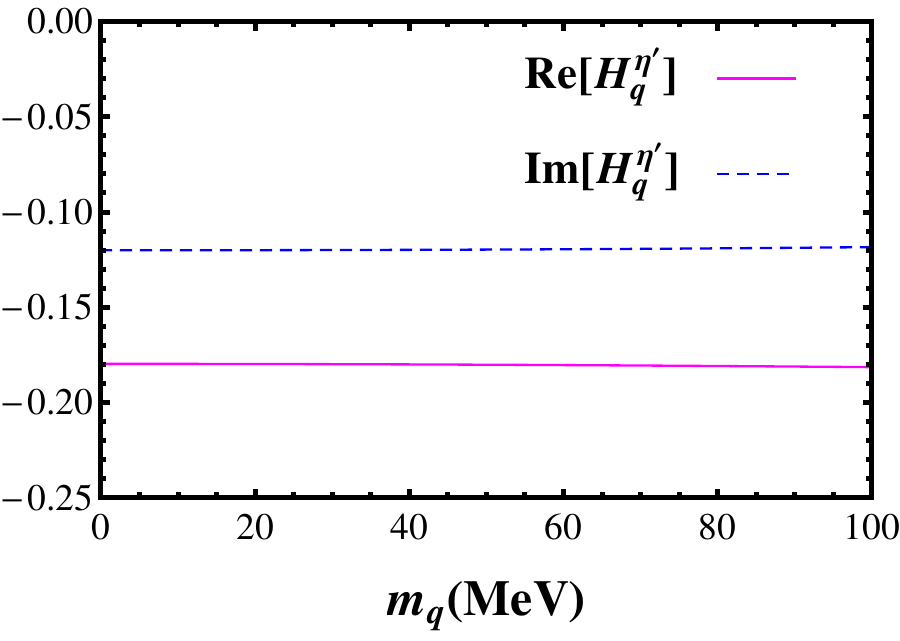}
\caption{\label{hmq}The $m_{q}$ dependence of real and imaginary parts of the dimensionless functions $H_{q}^{\eta}$ and $H_{q}^{\eta^{\prime}}$.}
\end{figure}

With algebraic identities
\begin{eqnarray}
&&q^{2}=2 \left(D_{4}+D_{5}-\frac{m^{2}}{2}\right)=2 \left(D_{1}+D_{2}+\frac{M^{2}}{2}\right),\nonumber\\
&&k\cdot p q^{2}-k\cdot q p\cdot q=\frac{2k\cdot p m^{2}}{M^{2}+m^{2}}\left(D_{1}+D_{2}\right)+\frac{2k\cdot p M^{2}}{M^{2}+m^{2}}(D_{4}+D_{5})
-k\cdot q(D_{4}-D_{5}),~~~~~~~
\end{eqnarray}
the loop function $I(u)$ can be decomposed into a sum of four- and three-point one-loop integrals
\begin{eqnarray}
 I(u)&=&\frac{1}{i \pi^{2}}\int\frac{\mathrm{d}^{4}q}{16}\Bigg{(}\frac{m^{4}(M^{2}-m^{2})}{M^{2}+m^{2}}\frac{1}{D_{2}D_{3}D_{4}D_{5}}
 -\frac{M^{2}(M^{2}-m^{2})^{2}}{2(M^{2}+m^{2})}\frac{1}{D_{1}D_{2}D_{3}D_{4}}\nonumber\\
 & &-\frac{3(M^{2}-m^{2})+2 k\cdot q}{2}\frac{1}{D_{1}D_{3}D_{4}}-\frac{M^{2}-m^{2}+2k\cdot q}{2}\frac{1}{D_{2}D_{3}D_{4}}\Bigg{)}+(u\leftrightarrow\bar{u}),~~~~~~~
\end{eqnarray}
which can be analytically calculated with the technique proposed in Ref.~\cite{Denner:1991qq} or the computer program $Package-\mathrm{X}$~\cite{Patel:2015tea,Patel:2016fam}.
Performing the convolution integral between the loop function $I(u)$ and the DA $\phi^{q}(u)$, we find that the function $H_{q}$ in Eq.~(\ref{scahq}) is very insensitive to the light quark masses. Specifically, the change of the absolute value of the function $H_{q}$ does not exceed 3\% when the value of $m_{q}$ goes from 0 to 100 MeV for all the three kinds of $\eta^{(\prime)}$ DAs in Fig.~\ref{DAs}. Actually, when the light quark propagator is near its mass shell, i.e., $q\sim\pm(u-\bar{u}) p$, the factor $(k\cdot p q^{2}-k\cdot q p\cdot q)$ in the numerator of
Eq.~(\ref{Iu}) tends to zero and cancels the light quark propagator pole.
In Fig.~\ref{hmq}, we show the dependence on $m_{q}$ of the functions $H_{q}^{\eta}=H_{q}{\mid}_{m=m_{\eta}}$ and $H_{q}^{\eta^{\prime}}=H_{q}{\mid}_{m=m_{\eta^{\prime}}}$ with a asymptotic DA. It is noticed that the one-loop QCD contribution to the decay $J/\psi\rightarrow\gamma\pi^{0}$ vanishes as a consequence of the antisymmetrical flavor wave function of $\pi^{0}$, even for $m_{u}\neq m_{d}$, which disagrees with the result in Ref.~\cite{Li:2005ug}.

\begin{figure}[!!htb]
\centering
\includegraphics[width=0.47\textwidth]{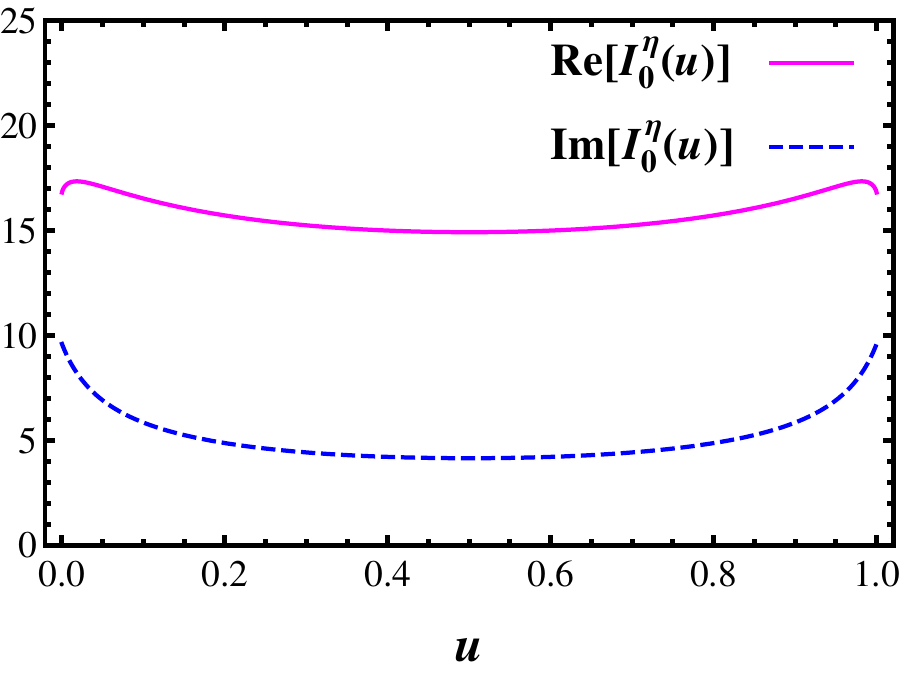}\hspace{0.5cm}
\includegraphics[width=0.47\textwidth]{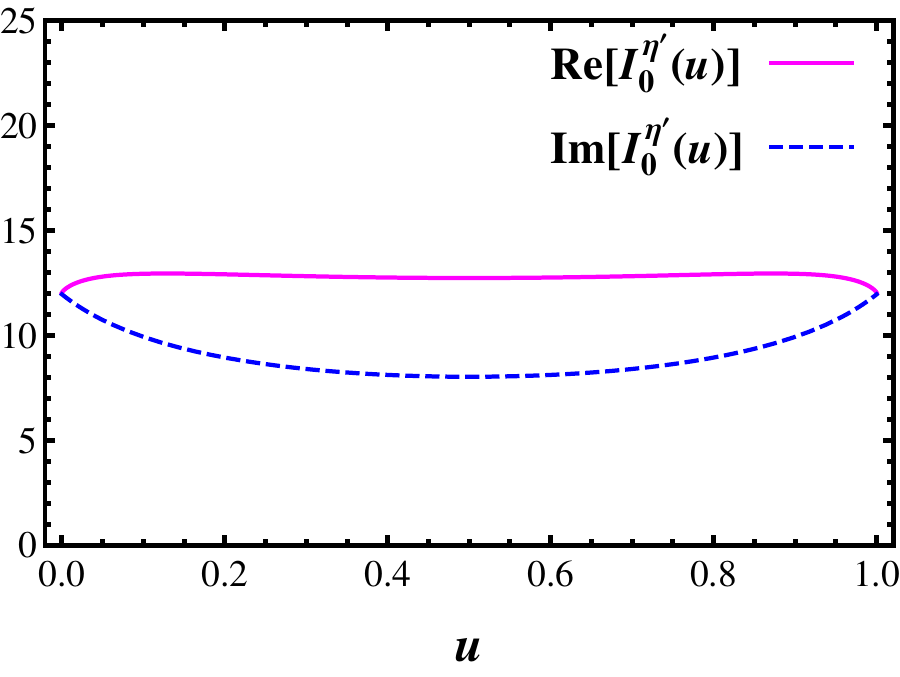}
\caption{\label{int0}The $u$ dependence of real and imaginary parts of the loop functions $I^{\eta}_{0}(u)$ and $I^{\eta^{\prime}}_{0}(u)$.}
\end{figure}

For simplicity, we can take the following limit safely
\begin{eqnarray}
I_{0}(u)&=&\lim_{m_{q}\to 0} I(u),\nonumber\\
H_{0}&=&-\frac{1}{16\pi^{2}}\frac{2}{1-x}\int\mathrm{d}u\phi^{q}(u)I_{0}(u),
\end{eqnarray}
that is $H_{q}(q=u,d,s)= H_{0}$. The expression of $I_{0}(u)$ is presented in the Appendix. Numerically,
we find that the loop function $I_{0}(u)$ is quite steady over the most region of $u$. In Fig.~\ref{int0}, we show the $u$ dependence of $I^{\eta}_{0}(u)=I_{0}(u){\mid}_{m=m_{\eta}}$ and $I^{\eta^{\prime}}_{0}(u)=I_{0}(u){\mid}_{m=m_{\eta^{\prime}}}$ with the range $u\in(0,1)$.
Unlike the result in the limit of $m^{2}/M^{2}\rightarrow0$~\cite{Kuhn:1983yr}, the loop functions $I^{\eta}_{0}(u)$ and $I^{\eta^{\prime}}_{0}(u)$ change slowly over the momentum fraction $u$ near the endpoints. As a consequence, the convolution integral between the loop function $I_{0}(u)$ and the DA becomes insensitive to the shapes of $\eta^{(\prime)}$ DAs. For example, the difference between the convolution integral of $I_{0}(u)$ with a ``narrow" DA (model I in Fig.~\ref{DAs}) and the one with a ``broad" DA (model II and III in Fig. \ref{DAs}) is less than $2\%$.

After these analyses, we return to the remaining calculations. The helicity amplitude $H_{QCD}^{q}$ in Eq.~(\ref{scalar}) can be simplified to
\begin{eqnarray}\label{scalarh2}
H_{QCD}^{q}&=&\frac{2Q_{c}}{9}\sqrt{4\pi\alpha_{e}}(4\pi\alpha_{s})^{2}\sqrt{\frac{3}{\pi M}}R_{\psi}(0)\frac{f_{\eta^{(\prime)}}}{M}H_{0}
\end{eqnarray}
with the effective decay constants
\begin{eqnarray}
f_{\eta^{\prime}}=f_{\eta^{\prime}}^{u}+f_{\eta^{\prime}}^{d}+f_{\eta^{\prime}}^{s},~~~~~~~
f_{\eta}=f_{\eta}^{u}+f_{\eta}^{d}+f_{\eta}^{s}.
\end{eqnarray}

\begin{figure}[!!htb]
\centering
\includegraphics[width=0.85\textwidth]{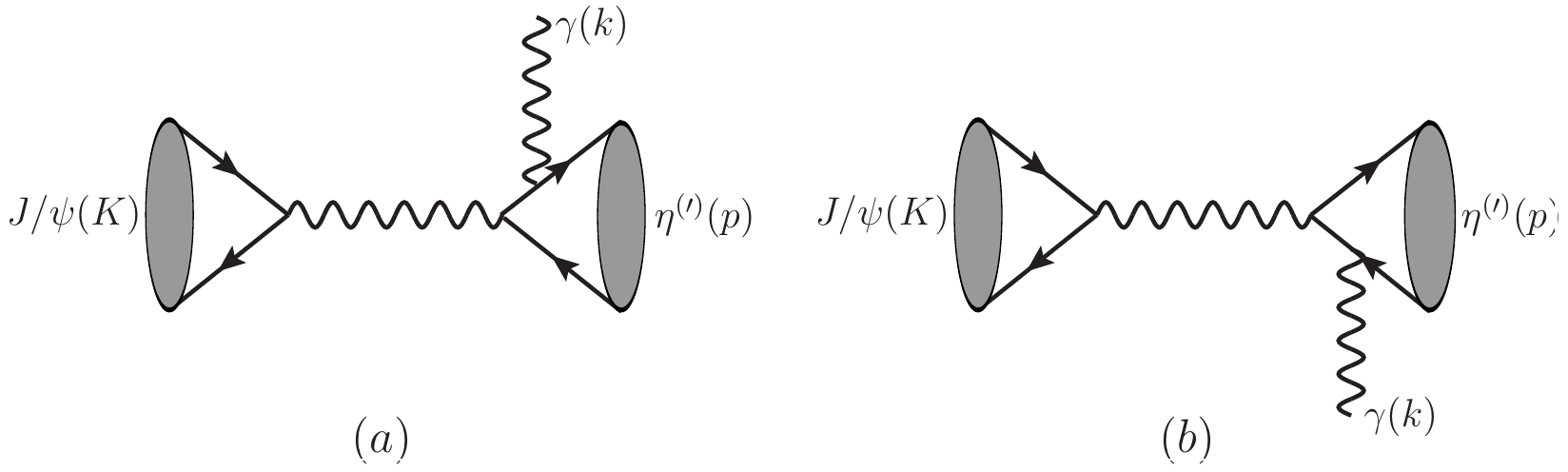}
\caption{\label{hqed}Feynman diagrams for the QED processes $J/\psi\rightarrow\gamma^{\ast}\rightarrow\gamma\eta^{(\prime)}$.}
\end{figure}

Besides the one-loop QCD contributions, QED processes $J/\psi\rightarrow\gamma^{\ast}\rightarrow\gamma\eta^{(\prime)}$ can also contribute to the decays $J/\psi\rightarrow\gamma \eta^{(\prime)}$. The corresponding Feynman diagrams are depicted in Fig.~\ref{hqed},
and the contribution reads
\begin{eqnarray}
H_{QED}&=&-Q_{c}(4\pi\alpha_{e})^{\frac{3}{2}}\sqrt{\frac{3}{\pi M}}R_{\psi}(0)
\sum\limits_{q=u,d,s}\frac{Q_{q}^{2}f_{\eta^{(\prime)}}^{q}}{M}h_{q},
\end{eqnarray}
where the dimensionless function is
\begin{eqnarray}
h_{q}&=&\frac{1-x}{2}\int\mathrm{d}u\phi^{q}(u)\left(\frac{1}{u-u\bar{u}x
-\frac{m_{q}^{2}}{M^{2}}+i\epsilon}+(u\leftrightarrow\bar{u})\right)
\end{eqnarray}
and the $Q_{q}$ represents the light quark charge.

\subsection{The contributions of the gluonic content of $\eta^{(\prime)}$}
\begin{figure}[!!htb]
\centering
\includegraphics[width=0.45\textwidth]{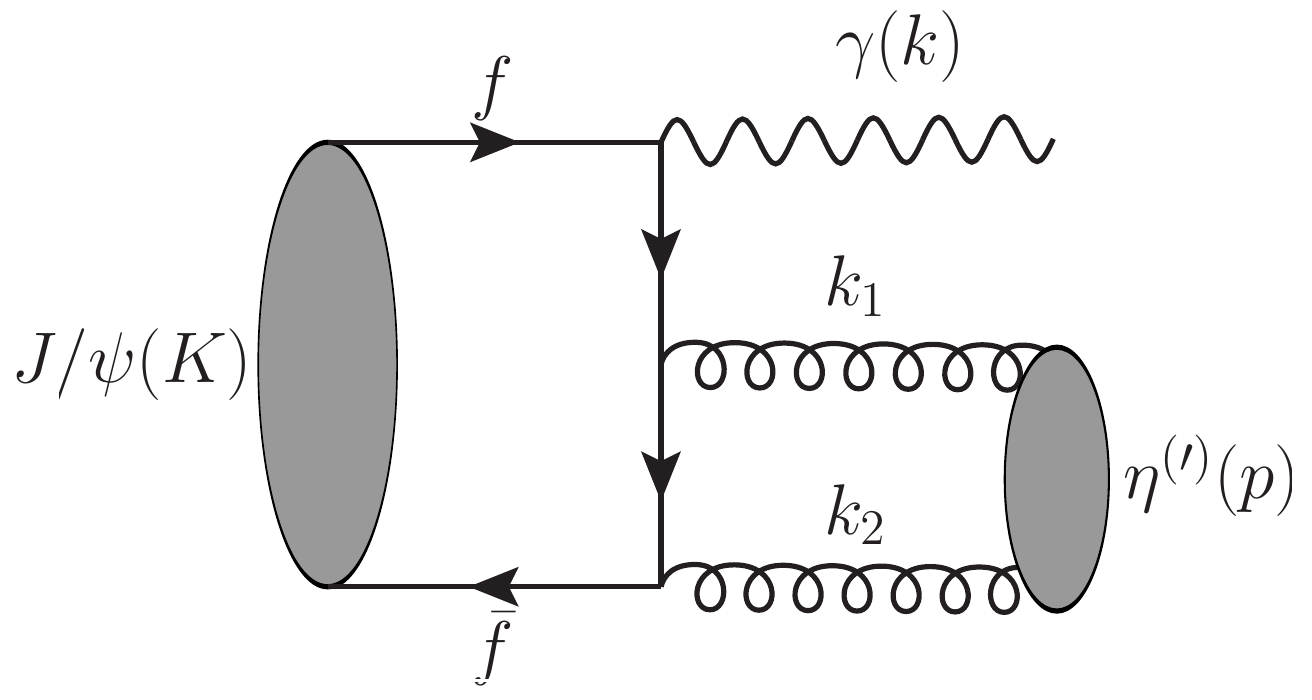}
\caption{\label{hQCDg}One typical Feynman diagram for $J/\psi\rightarrow \gamma \eta^{(\prime)}$ with the
 gluonic content of $\eta^{(\prime)}$.}
\end{figure}

The gluonic content of $\eta^{(\prime)}$ can contribute to the $J/\psi\rightarrow\gamma\eta^{(\prime)}$ at the tree level. However, such contributions are suppressed by a factor of $m^{2}/M^{2}$~\cite{Ma:2002ww,Li:2007dq}. The corresponding Feynman diagram is shown in Fig.~\ref{hQCDg}. There are other two Feynman diagrams from permutations of the photon and the gluon legs.
The leading twist in the light-cone expansion of the matrix elements of the meson $\eta^{(\prime)}$ over two-gluon fields is~\cite{Kroll:2002nt,Ball:2007hb,Agaev:2014wna}:
\begin{eqnarray}
\langle\eta^{(\prime)}(p)|A_{\alpha}^{a}(x)A_{\beta}^{b}(y)|0\rangle&=&\frac{1}{4}\epsilon_{\alpha\beta\rho\sigma}
\frac{k^{\rho}p^{\sigma}}{p\cdot k}\frac{C_{F}}{\sqrt{3}}\frac{\delta^{ab}}{8}f_{\eta^{(\prime)}}^{g}\int\mathrm{d}u e^{i(up\cdot x+\bar{u}p\cdot y)}\frac{\phi^{g}(u)}{u(1-u)}
\end{eqnarray}
with the effective decay constant $f_{\eta^{(\prime)}}^{g}=\frac{1}{\sqrt{3}}\left(f_{\eta^{(\prime)}}^{u}+f_{\eta^{(\prime)}}^{d}+f_{\eta^{(\prime)}}^{s}\right)$ and the gluonic twist-2 DA~\cite{Agaev:2014wna,Ball:2007hb,Alte:2015dpo}
\begin{eqnarray}
\phi^{g}(u)&=&30u^{2}(1-u)^{2}\sum_{n=2,4\cdots}c^{g}_{n}(\mu)C_{n-1}^{\frac{5}{2}}(2u-1).
\end{eqnarray}
The corresponding contributions can be expressed as
\begin{eqnarray}
H_{QCD}^{g}&=&\frac{2Q_{c}}{9}\sqrt{4\pi\alpha_{e}}(4\pi\alpha_{s})\frac{R_{\psi}(0)}{\sqrt{\pi M}} \frac{f_{\eta^{(\prime)}}^{g}}{M}H_{g}
\end{eqnarray}
with
\begin{eqnarray}\label{hg}
H_{g}=\int\mathrm{d}u \frac{\phi_{g}(u)}{u(1-u)}\frac{2 x(2 u-1)}{1-x^{2}(1-2 u)^{2}}.
\end{eqnarray}
From the Eq.~(\ref{hg}), one can find that $H_{g}$ is proportional to a suppression factor of $x=m^{2}/M^{2}$. Actually, the leading twist gluonic content contributions are almost two on-shell gluons contributions, which are suppressed by the factor $m^{2}/M^{2}$ due to the special form of the Ore-Powell matrix elements as found in Refs.~\cite{Krammer:1978qp,Billoire:1978xt} years ago.

\section{Numerical results}
\label{sec:numerical analysis}

The decay widths of $J/\psi\rightarrow \gamma\eta^{(\prime)}$ can be expressed as
\begin{eqnarray}
\Gamma(J/\psi\rightarrow \gamma\eta^{(\prime)})=\frac{2}{3}\frac{1}{16\pi}\frac{1-x}{M}{\mid}H_{QCD}^{q}+H_{QCD}^{g}+H_{QED}{\mid}^{2}.
\end{eqnarray}
In order to remove the uncertainties from $R_{\psi}(0)$, we relate the decay widths $\Gamma(J/\psi\rightarrow \gamma\eta^{(\prime)})$ to $\Gamma(J/\psi\rightarrow e^{+}e^{-})$
\begin{eqnarray}
\mathcal{B}(J/\psi\rightarrow \gamma\eta^{(\prime)})=\frac{\Gamma(J/\psi\rightarrow \gamma\eta^{(\prime)})}
{\Gamma(J/\psi\rightarrow e^{+}e^{-})}\mathcal{B}^{exp}(J/\psi\rightarrow e^{+}e^{-})
\end{eqnarray}
with the leptonic decay width~\cite{Mackenzie:1981sf}
\begin{eqnarray}
\Gamma(J/\psi\rightarrow e^{+}e^{-})=\frac{4\alpha_{e}^{2}Q_{c}^{2}}{M^{2}}\big{|}R_{\psi}(0)\big{|}^{2}\left(1-\frac{16}{3}\alpha_{s}\right)
\end{eqnarray}
and its experimental value~\cite{Tanabashi:2018oca}
\begin{eqnarray}
\mathcal{B}^{exp}(J/\psi\rightarrow e^{+}e^{-})=(5.971\pm0.032)\times 10^{-2}.
\end{eqnarray}
For numerical calculations, all the values of meson masses, $\Gamma_{J/\psi}$ and $f_{\pi}$ are quoted from the PDG~\cite{Tanabashi:2018oca}. We use the FKS scheme for the $\eta-\eta^{\prime}$ mixing~\cite{Feldmann:1998vh}, and then the effective decay constants $f_{\eta^{(\prime)}}^{q}$ can be parameterized as
\begin{eqnarray}
f_{\eta}^{u(d)}&=&\frac{f_{q}}{\sqrt{2}}\cos\phi,~~~~~~~   f_{\eta}^{s}=-f_{s}\sin\phi, \nonumber\\
f_{\eta^{\prime}}^{u(d)}&=&\frac{f_{q}}{\sqrt{2}}\sin\phi,~~~~~~~   f_{\eta^{\prime}}^{s}=f_{s}\cos\phi.
\end{eqnarray}
For the three phenomenological parameters, i.e., the mixing angle $\phi$ and the decay constants $f_{q}$, $f_{s}$, they have been determined in different methods~\cite{Feldmann:1998vh,Escribano:2005qq,Escribano:2007cd,Cao:2012nj,Escribano:2013kba}. Here we take the up-to-date values from Refs.~\cite{Escribano:2005qq,Escribano:2013kba} which are recapitulated in Table~\ref{tab:parameters}.
\begin{table}[!!htb]
  \caption{\label{tab:parameters} The values of $\phi$, $f_{q}$ and $f_{s}$ obtained with three phenomenological approaches~\cite{Escribano:2005qq,Escribano:2013kba}.}
\vspace{0.2cm}
\centering
  \begin{tabular}{lccc}
  \hline\hline
                         ~~~~~&~~~~~$\phi^{\circ}$~~~~~&~~~~~$f_{q}/f_{\pi}$~~~~~&~~~~~$f_{s}/f_{\pi}$~~~~~ \\
  \hline
  LEPs~\cite{Escribano:2005qq}   & $40.6\pm0.9$& $1.10\pm0.03$  & $1.66\pm0.06$ \\
  $\eta$TFF~\cite{Escribano:2013kba}& $40.3\pm1.8$& $1.06\pm0.01$  & $1.56\pm0.24$ \\
  $\eta^{\prime}$TFF~\cite{Escribano:2013kba}    & $33.5\pm0.9$& $1.09\pm0.02$  & $0.96\pm0.04$ \\
  \hline\hline
  \end{tabular}
\end{table}
The values in the first line are extracted from the low energy processes(LEPs) $V\rightarrow\eta^{(\prime)}\gamma$, $\eta^{(\prime)}\rightarrow V\gamma$($V=\rho,\omega,\phi$). The second and the third lines are the results extracted with rational approximations for the $\eta^{(\prime)}$ transition form factor(TFF) $F_{\gamma^{\ast}\gamma\eta^{(\prime)}}(Q^{2})$. The results of both the first and the second lines are generally consistent with the known FKS results~\cite{Feldmann:1998vh} where the ratio $R_{J/\psi}$ with the approximation of nonperturbative matrix elements $\langle0|G_{\mu\nu}^a\tilde{G}^{a,\mu\nu}|\eta^{(\prime)}\rangle$ was adopted. While in the third line, the parameters are extracted with rational approximations for the $\eta^{\prime}$ TFF $F_{\gamma^{\ast}\gamma\eta^{\prime}}(Q^{2})$, which is in accord with the BaBar measurement in the timelike region at $q^{2}=112$ GeV$^{2}$~\cite{Aubert:2006cy}.

The Gegenbauer moments $c^{q}_{2}(\mu)$, $c^{q}_{4}(\mu)$ still have large uncertainty as depicted in Table~\ref{tab:coefficients}. Fortunately, the dimensionless function $H_{q}$ in Eq.~(\ref{scahq}) is insensitive to the shapes of the $\eta^{(\prime)}$ DAs as we have shown in section~\ref{subsec:QCDq}. Furthermore, both $H_{g}$ and $h_{q}$ are an order of magnitude smaller than the $H_{q}$, therefore, the uncertainty of the Gegenbauer moments impacts our numerical calculations lightly (less than 2\% among the results with three different models in Table~\ref{tab:coefficients}). So in the following numerical calculations, we choose the Model I with
\begin{eqnarray}
c^{q}_{2}(\mu_{0})=0.10,~~~~~~~ c^{q}_{4}(\mu_{0})=0.10,~~~~~~~
c_{2}^{g}(\mu_{0})=-0.26.
\end{eqnarray}

As known, it is very hard to give precise predictions for individual decay width. With the nonperturbative matrix
elements $\langle0|G_{\mu\nu}^a\tilde{G}^{a,\mu\nu}|\eta^{(\prime)}\rangle$, the decay widths of $J/\psi\rightarrow\gamma\eta^{(\prime)}$ have been given in Ref. \cite{Novikov:1979uy}
\begin{eqnarray}\label{npe}
\Gamma(J/\psi\rightarrow\gamma\eta^{(\prime)})&=&\frac{2^{3}}{5^{2}3^{10}}\frac{\alpha_{e}^{3}\alpha_{s}^{2}}{\pi}
\left(\frac{M^{4}}{m_{c}^{8}}\right)
\left(1-\frac{m_{\eta^{(\prime)}}^{2}}{{M}^{2}}\right)^{3}
\frac{\mid\langle0|G_{\mu\nu}^a\tilde{G}^{a,\mu\nu}|\eta^{(\prime)}\rangle\mid^{2}}
{\Gamma(J/\psi\rightarrow e^{+}e^{-})},
\end{eqnarray}
where the factor $m_{c}^{-8}$ will bring very large uncertainty. While, in the pQCD approach employed in this paper, there
also exist large uncertainty due to the factor $\alpha_{s}^{4}(\mu)$.

For comparison, in Table~\ref{tab:nonper}, we present results of these two methods with $m_{c}=1.5~\mathrm{GeV}$ and
$\alpha_{s}=0.34$ as benchmarks. For the results in the second column, the matrix elements $\langle0|G_{\mu\nu}^a\tilde{G}^{a,\mu\nu}|\eta^{(\prime)}\rangle$ are evaluated with the updated values of $\phi$, $f_{q}$ and $f_{s}$ in the
first line of Table~\ref{tab:parameters}. Generally, one may expect the order of magnitude could be correctly predicted
by the both approaches. However, we find the pQCD estimation of $\mathcal{B}(J/\psi\rightarrow\gamma\eta)$ is too small to
be comparable to its experimental one. The reason may be due to our choice of the inputs $\phi$, $f_{q}$ and $f_{s}$ extracted
from low energy processes, or our understanding of $\eta-\eta^{\prime}$ mixing scheme is incomplete which is beyond the scope
of this paper.

 \begin{table}[!!htb]
  \caption{\label{tab:nonper}The branching ratios $\mathcal{B}(J/\psi\rightarrow\gamma\eta^{(\prime)})$ obtained with nonperturbative gluonic matrix elements and pQCD approaches.}
  \vspace{0.2cm}
  \centering
  \begin{tabular}{lccc}
  \hline\hline
              ~~~~~&~~~~~$\langle0|G_{\mu\nu}^a\tilde{G}^{a,\mu\nu}|\eta^{(\prime)}\rangle$~\cite{Novikov:1979uy}~~~~~&~~~~~pQCD
              ~~~~~&~~~~~Exp.~\cite{Ablikim:2005je,Pedlar:2009aa,Ablikim:2010kp,Tanabashi:2018oca}~~~~~  \\
  \hline
$\mathcal{B}(J/\psi\rightarrow\gamma\eta^{\prime})$&$1.96\times10^{-3}\left(\frac{1.5~\mathrm{ GeV}}{m_{c}}\right)^{8}$
&$4.91\times10^{-3}\left(\frac{\alpha_{s}(\mu)}{0.34}\right)^{4}$&(5.13$\pm$0.17)$\times10^{-3}$\\
$\mathcal{B}(J/\psi\rightarrow\gamma\eta)$&$3.99\times10^{-4}\left(\frac{1.5~\mathrm{ GeV}}{m_{c}}\right)^{8}$
&$9.84\times10^{-6}\left(\frac{\alpha_{s}(\mu)}{0.34}\right)^{4}$&(11.04$\pm$0.34)$\times10^{-4}$\\
  \hline\hline
  \end{tabular}
\end{table}

Recently, the BaBar collaboration~\cite{Aubert:2006cy} has made the measurements of the $\eta$ and $\eta^{\prime}$
TFFs at $q^{2}=112~\mathrm{GeV}^{2}$, which have challenged the theoretical prediction very much.
It is true that precise theoretical estimation of the TFFs is hard, due to uncertainties in the $\phi$, $f_{q}$,
$f_{s}$, the DAs of $\eta^{(\prime)}$ and even the mixing scheme at high energy. In the literature, there are extensive discussions of this hot topic~\cite{Agaev:2014wna}. In Ref.~\cite{Escribano:2013kba}, Escribano {\it et al.} have presented a treatment of $\eta$ and $\eta^{\prime}$ TFFs
with rational approximations and extracted these parameters in Table~\ref{tab:parameters}. With these inputs and $\alpha_{s}=0.34$, we show our results in Table~\ref{tab:QCDq}. From the Table, we find $R_{J/\psi}$ is in good agreement with its experimental value $R_{J/\psi}^{exp}=4.65\pm0.21$~\cite{Tanabashi:2018oca}, only when we adopt the set of parameter values of $\eta^{\prime}$TFF.
For other two sets of parameter values, $\mathcal{B}(J/\psi\rightarrow\gamma\eta)$ is estimated to be too small, which results in
$R_{J/\psi}$ two orders higher than $R_{J/\psi}^{exp}$.

\begin{table}[!!htb]
  \caption{\label{tab:QCDq}The branching ratios $\mathcal{B}(J/\psi\rightarrow\gamma\eta^{(\prime)})$ with three sets
  of inputs obtained in Refs.~\cite{Escribano:2005qq,Escribano:2013kba}.}
  \vspace{0.2cm}
  \centering
  \begin{tabular}{lcccc}
  \hline\hline
              ~~~~~&~~~~~LEPs~~~~~&~~~~~$\eta$TFF~~~~~&~~~~~$\eta^{\prime}$TFF~~~~~&~~~~~Exp.~\cite{Ablikim:2005je,Pedlar:2009aa,Ablikim:2010kp,Tanabashi:2018oca}~~~~~  \\
  \hline
 $\mathcal{B}(J/\psi\rightarrow\gamma\eta^{\prime})$  & $4.91\times10^{-3}$ &$4.43\times10^{-3}$&$2.59\times10^{-3}$&$(5.13\pm0.17)\times10^{-3}$\\
 $\mathcal{B}(J/\psi\rightarrow\gamma\eta)$&$9.84\times10^{-6}$&$1.74\times10^{-5}$&$5.52\times10^{-4}$
 &$(11.04\pm0.34)\times10^{-4}$\\
 $R_{J/\psi}$               &$499.19$ &$254.18$&$4.70$ &$4.65\pm0.21$  \\
  \hline\hline
  \end{tabular}
\end{table}

In Table~\ref{tab:total}, we present results with the contributions due to the gluonic content of $\eta^{(\prime)}$ and the one from QED processes $J/\psi\rightarrow\gamma^{\ast}\rightarrow\gamma\eta^{(\prime)}$. We can find that such contributions enhance
$\mathcal{B}(J/\psi\rightarrow\gamma\eta)$ much. However, $\mathcal{B}(J/\psi\rightarrow\gamma\eta)$ is still far from its
experimental value for both LEPs and $\eta$TFF values of $\phi$, $f_{q}$ and $f_{s}$. Only $\eta^{\prime}$TFF set of parameter
values can give $\mathcal{B}(J/\psi\rightarrow\gamma\eta^{(\prime)})$ and $R_{J/\psi}$ comparable with their experimental data.

\begin{table}[!!htb]
  \caption{\label{tab:total}The same as the caption of Table~\ref{tab:QCDq}, but including contributions of the QED and the gluonic content of $\eta^{(\prime)}$.}
  \vspace{0.2cm}
  \centering
  \begin{tabular}{lcccc}
  \hline\hline
          ~~~~~&~~~~~LEPs    ~~~~~&~~~~~$\eta$TFF~~~~~&~~~~~$\eta^{\prime}$TFF~~~~~&~~~~~Exp.~\cite{Ablikim:2005je,Pedlar:2009aa,Ablikim:2010kp,Tanabashi:2018oca}~~~~~ \\
  \hline
  $\mathcal{B}(J/\psi\rightarrow\gamma\eta^{\prime})$&$6.01\times10^{-3}$&$5.43\times10^{-3}$&$3.19\times10^{-3}$ &(5.13$\pm$0.17)$\times10^{-3}$  \\
  $\mathcal{B}(J/\psi\rightarrow\gamma\eta)$  &$3.02\times10^{-5}$   &$4.25\times10^{-5}$&$7.40\times10^{-4}$  &(11.04$\pm$0.34)$\times10^{-4}$ \\
  $R_{J/\psi}$       &$198.91$   &$127.60$  &$4.31$&$4.65\pm0.21$  \\
  \hline\hline
  \end{tabular}
\end{table}

\begin{figure}[!!htb]
\centering
      \includegraphics[width=0.6\textwidth]{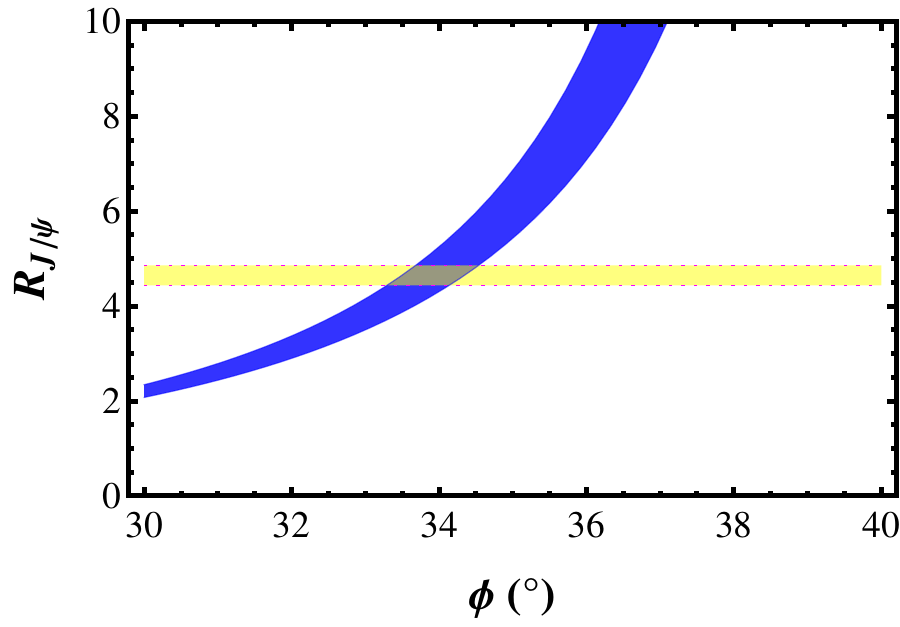}
\caption{\label{Rjpsirr}The dependence of $R_{J/\psi}$ on the mixing angle $\phi$. The blue band is our calculated results with the uncertainties of the $\Gamma^{exp}(\eta^{(\prime)}\rightarrow\gamma\gamma)$ included. The hatched band indicates the experimental value of $R_{J/\psi}$ with $1 \sigma$ uncertainty.}
\end{figure}

In the remaining part of this section, we would present a determination of $\phi$ without the input values in Table~\ref{tab:parameters}. The ratio $R_{J/\psi}$ in our calculation is
\begin{eqnarray}
R_{J/\psi}=\frac{M^{2}-m_{\eta^{\prime}}^{2}}{M^{2}-m_{\eta}^{2}}
\frac{{\mid}H_{QCD}^{q}+H_{QCD}^{g}+H_{QED}{\mid}^{2}_{m=m_{\eta^{\prime}}}}
{{\mid}H_{QCD}^{q}+H_{QCD}^{g}+H_{QED}{\mid}^{2}_{m=m_{\eta}}}.
\end{eqnarray}
Since the scalar functions are insensitive to the shapes of the $\eta^{(\prime)}$ DAs, the ratio $R_{J/\psi}$ mainly depends on the angle $\phi$ and the ratio $f_{s}/f_{q}$. However, with the help of the ratio
\begin{eqnarray}
\frac{\Gamma(\eta\rightarrow\gamma\gamma)}{\Gamma(\eta^{\prime}\rightarrow\gamma\gamma)}
=\frac{m_{\eta}^{3}}{m_{\eta^{\prime}}^{3}}\left(\frac{5\sqrt{2} \frac{f_{s}}{f_{q}}-2 \tan \phi }{5\sqrt{2} \frac{f_{s}}{f_{q}} \tan \phi +2 }\right)^{2}
\end{eqnarray}
and the experimental values~\cite{Babusci:2012ik,Tanabashi:2018oca}
\begin{eqnarray}
\Gamma^{exp}(\eta^{\prime}\rightarrow\gamma\gamma)=4.36(14)~\mathrm{KeV},\quad\quad
\Gamma^{exp}(\eta\rightarrow\gamma\gamma)=0.516(18)~\mathrm{KeV},
\end{eqnarray}
the ratio $R_{J/\psi}$ becomes a function which only depends on the mixing angle $\phi$. In Fig.~\ref{Rjpsirr}, we show
the dependence of the ratio $R_{J/\psi}$ on the mixing angle $\phi$.
As displayed by the horizontal dashed lines in Fig.~\ref{Rjpsirr} of the experimental measurement $R^{exp}_{J/\psi}=4.65\pm0.21$~\cite{Tanabashi:2018oca}, one can find
\begin{eqnarray}
\phi&=&33.9^{\circ}\pm0.6^{\circ},
\end{eqnarray}
which is in good agreement with $\eta^{\prime}$TFF result $\phi=33.5^{\circ}\pm0.9^{\circ}$~\cite{Escribano:2013kba}\renewcommand{\thefootnote}{\fnsymbol{footnote}}\footnote[2]{It is noticed that the predicted $\eta$ TFF is $\lim_{Q^{2}\rightarrow+\infty}Q^{2}F_{\eta\gamma^{\ast}\gamma}(Q^{2})=(0.160\pm0.024)~\mathrm{GeV}$, which is not in line with the BaBar measurement $q^{2}{\mid}F_{\eta\gamma^{\ast}\gamma}(q^{2}){\mid}_{q^{2}=112~\mathrm{GeV}^{2}}=(0.229\pm0.031)~\mathrm{GeV}$, while the predicted $\eta^{\prime}$ TFF, $\lim_{Q^{2}\rightarrow+\infty}Q^{2}F_{\eta^{\prime}\gamma^{\ast}\gamma}(Q^{2})=(0.255\pm0.004)~\mathrm{GeV}$, is in accord
with the BaBar measurement $q^{2}{\mid}F_{\eta^{\prime}\gamma^{\ast}\gamma}(q^{2}){\mid}_{q^{2}=112~\mathrm{GeV}^{2}}=(0.251\pm0.021)~\mathrm{GeV}$.
More discussions could be found in Ref.~\cite{Escribano:2013kba}.}, but in clear disagreement with the $\phi=40.6^{\circ}\pm0.9^{\circ}$~\cite{Escribano:2005qq} extracted from low energy processes
with nonperturbative methods. It is noticed that the lattice calculation of the UKQCD collaboration~\cite{Gregory:2011sg} indicates a value $\phi^{fit}\sim34^{\circ}$, while the ETM collaboration~\cite{Ottnad:2012fv,Michael:2013gka,Ottnad:2017bjt} gives $\phi$ in the range of $40^{\circ}\sim46^{\circ}$.

Last but not least, we would compare our result of the loop function with the one obtained by K\"{u}hn~\cite{Kuhn:1983yr}.
Since, besides the approximation $M\approx2 m_{c}$, no further approximation is made, so our loop function $I_{0}(u)$, as shown
in the Appendix, is much more complicated than the one in Ref.~\cite{Kuhn:1983yr}. However, in the limit of $m^{2}/M^{2}\rightarrow0$, our $I_{0}(u)$ is reduced to
the weight function $\mathcal{W}(u)$ in the Eq.~(11) of Ref.~\cite{Kuhn:1983yr}.

\section{Summary}
\label{sec:conclusion}
In this paper, we have revisited the radiative decays $J/\psi\rightarrow \gamma\eta^{(\prime)}$ in detail in the framework of pQCD. Comparing to the pioneer work~\cite{Korner:1982vg}, we do not take the weak-binding approximation for the final mesons and evaluate the involved one-loop integrals analytically with the light quark masses kept. Moreover, we also consider the contributions from the QED processes and the gluonic contents of $\eta^{(\prime)}$.
Different from the results obtained in Ref.~\cite{Li:2005ug}, our numerical results are insensitive to the light quark masses.
In addition, we find that the $\mathcal{B}(J/\psi\rightarrow\gamma\eta^{(\prime)})$ are insensitive to the shapes of the $\eta^{(\prime)}$ DAs.

Using three sets of values for $\phi$, $f_{q}$ and $f_{s}$, namely LEPs~\cite{Escribano:2005qq}, $\eta$TFF~\cite{Escribano:2013kba} and $\eta^{\prime}$TFF~\cite{Escribano:2013kba}, we have presented our numerical
results in Table~\ref{tab:QCDq} and~\ref{tab:total}, where $\mathcal{B}(J/\psi\rightarrow\gamma\eta)$ is too
small to be comparable with the experimental one for the parameters extracted from LEPs and $\eta$ TFF. Only with the values extracted from $\eta^{\prime}$ TFF, which is in accord with the BaBar measurement in the timelike region at $q^{2}=112$ GeV$^{2}$~\cite{Aubert:2006cy}, our results of $\mathcal{B}(J/\psi\rightarrow\gamma\eta^{(\prime)})$ and $R_{J/\psi}$ are
in good agreement with the experimental data.

As a crossing check, we use our calculation of $R_{J/\psi}$ and $\Gamma^{exp}(\eta^{(\prime)}\rightarrow\gamma\gamma)$
as inputs to extract the value of $\phi$, and find $\phi=33.9^{\circ}\pm0.6^{\circ}$, which is in good agreement with the $\eta^{\prime}$TFF one $\phi=33.5^{\circ}\pm0.9^{\circ}$ remarkably. However, such a small $\phi$ differs too much from
the $\phi=40.6^{\circ}\pm0.9^{\circ}$ extracted from low energy processes and $J/\psi\rightarrow\gamma\eta^{(\prime)}$ with
nonperturbative matrix elements $\langle0|G_{\mu\nu}^a\tilde{G}^{a,\mu\nu}|\eta^{(\prime)}\rangle$ due to $U_{A}(1)$ anomaly
dominance argument. The difference may arise from $g^{\ast}g^{\ast}-\eta^{(\prime)}$ TFF used in our calculation, in like manner,
$\gamma^{\ast}\gamma-\eta^{\prime}$ TFF in the extraction of $\phi$ from the BaBar measurement of $q^{2}{\mid}F_{\eta^{\prime}\gamma^{\ast}\gamma}(q^{2}){\mid}_{q^{2}=112~\mathrm{GeV}^{2}}$. Anyhow, the physics under the difference
is interesting and certainly worth further investigations.

\section*{Acknowledgements}
This work is supported by the National Natural Science Foundation of China under Grant Nos.~11675061, 11775092 and 11435003.

\appendix

\section*{Appendix: The dimensionless function $H_{0}$}
\label{sec:analytic results}
The dimensionless function $H_{0}$ reads
\begin{align*}
 H_{0}=-\frac{1}{16\pi^{2}}\frac{2}{1-x}\int^{1}_{0}\mathrm{d}u\phi^{q}(u)I_{0}(u).\tag{1}
\end{align*}
Since the $H_{0}$ is insensitive to the shapes of the $\eta^{(\prime)}$ DAs,
we simply take the asymptotic DA $\phi^{q}(u)=6 u (1-u)$ in the following discussion.
The analytical expression of the one-loop function $I_{0}(u)$ can be expressed as
\begin{align*}
I_{0}(u)= g_{1}(\xi)+g_{2}(\xi)+g_{3}(\xi)+g_{4}(\xi)+g_{5}(\xi)+g_{6}(\xi)+(\xi\rightarrow-\xi)\tag{2}
\end{align*}
with $\xi=1-2u$ and
\begin{align*}
g_{1}(\xi)=&\frac{2 x \left[1+3 (1-x) \xi-x \xi^2\right]}{(1-x) (1+\xi) (1-x \xi)}\bigg{[}\mathrm{Li}_{2}\left(\frac{1-x \xi}{2-x-x \xi}\right)+\mathrm{Li}_{2}\left(-\frac{(1-2 x) (1-x \xi)}{x (1+\xi-2 x \xi)}\right)\\
& +\mathrm{Li}_{2}\left(\frac{1+\xi+2 x^2 \xi (1+\xi)-x \left(1+4 \xi+\xi^2\right)}{x \xi (-2+x+x \xi)}-i\xi\epsilon\right)-\mathrm{Li}_{2}\left(\frac{1-x \xi}{1+\xi-2 x \xi}\right)\\
& -\mathrm{Li}_{2}\left(-\frac{(1-2 x) (2-x-x \xi)}{x (1+\xi-2 x \xi)}\right)-\mathrm{Li}_{2}\left(\frac{1+x^2 \xi (1+\xi)-x (1+2 \xi)}{x \xi (-2+x+x \xi)}-i\xi\epsilon\right)\\
& -\mathrm{Li}_{2}\left(\frac{1+\xi-2 x \xi}{2-x-x \xi}\right)+\mathrm{Li}_{2}\left(\frac{2-x-x \xi}{1+\xi-2 x \xi}\right)
\bigg{]},\\
g_{2}(\xi)=&-\frac{2 (1-x)}{(1+x) (1-\xi) (1-x \xi)}\bigg{[}-4 \mathrm{Li}_{2}(x)+2 \mathrm{Li}_{2}(2 x-1)-2 \mathrm{Li}_{2}\left(-\xi+x\xi+x \xi^{2}\right)\\
& +2 \mathrm{Li}_{2}\left(\xi-x \xi+x \xi^2\right)-\ln^2 x-2 i \pi \ln x+4 i \pi  \ln\left(\frac{2-x-x \xi}{1+\xi-2 x \xi}\right)\bigg{]},\\
g_{3}(\xi)=&\frac{2 x \left[(1+x^2 \xi) (3-\xi) -x \left(5-2 \xi+\xi^2\right)\right]}{(1-x^{2}) (1-\xi) (1-x \xi)}\bigg{[}-\mathrm{Li}_{2}\left(-\frac{(1-x (2-\xi)) \xi}{(1-x (1-\xi)) (1-\xi)}\right)\\
& -\mathrm{Li}_{2}\left(-\frac{1-x (2-\xi)}{x^2 (1-\xi)}\right)+\mathrm{Li}_{2}\left(\frac{x^2 \xi}{1-x (1-\xi)}\right)+2\mathrm{Li}_{2}\left(\frac{x (2-\xi)-1}{x (1-\xi)}\right)\bigg{]},\\
g_{4}(\xi)=&-\frac{8x}{(1+x) (1-\xi^{2})}\bigg{[}2 i \pi  \ln\left(\frac{2-x-x \xi}{1+\xi-2 x \xi}\right)-2 \mathrm{Li}_{2}\left(\frac{3x-1}{2 x}\right)+\mathrm{Li}_{2}\left(\frac{1-3 x}{2-4 x}\right)\\
& - \mathrm{Li}_{2}\left(\frac{-1+x (2+\xi)}{x^2 (1+\xi)}\right)- \mathrm{Li}_{2}\left(\frac{x^2}{2 x-1}\right)+ \mathrm{Li}_{2}\left(\frac{3 x-1}{2 x^2}\right)+ \mathrm{Li}_{2}\left(\frac{x^2 \xi}{-1+x+x \xi}\right)\\
& +2 \mathrm{Li}_{2}\left(\frac{-1+x (2+\xi)}{x (1+\xi)}\right)- \mathrm{Li}_{2}\left(\frac{\xi (1- 2x-x\xi)}{(1+\xi) (1-x-x \xi)}\right) \bigg{]},\\
g_{5}(\xi)=&\frac{1}{3(1-x) (1-\xi)}\bigg{[} (2- 3x+x\xi)\big{[}\pi^{2}-3 \ln^2 x-12\mathrm{Li}_{2}(x)-6 i \pi   \ln x \big{]}\nonumber\\
& +12 i\pi (2-x-x \xi) \ln\frac{2-x-x \xi}{1+\xi-2 x \xi}\bigg{]},\\
g_{6}(\xi)=&\frac{2(1+x\xi)\ln[(1-\xi)(1+x \xi)]}{1-x+x\xi}+\frac{4 (x-1)\ln(2-2x)}{(1-2x)(1-\xi)}\\
& +\frac{2(1+\xi)(1-x\xi)}{(1-\xi)(1-x-x\xi)}
\ln[(1-x\xi)(1+\xi)].\tag{3}
\end{align*}
In the above expressions, the dilogarithm function $\mathrm{Li}_{2}(x)$ and the logarithm function $\ln(x)$ without ``$i \epsilon$" prescription are defined as
\begin{align*}
\mathrm{Li}_{2}(x)&= \lim_{\epsilon\rightarrow 0^{+}}\mathrm{Li}_{2}(x-i \epsilon),\\
\ln(x)&= \lim_{\epsilon\rightarrow 0^{+}}\ln(x+i \epsilon).\tag{4}
\end{align*}

\begin{figure}[!!htb]
\centering
      \includegraphics[width=0.6\textwidth]{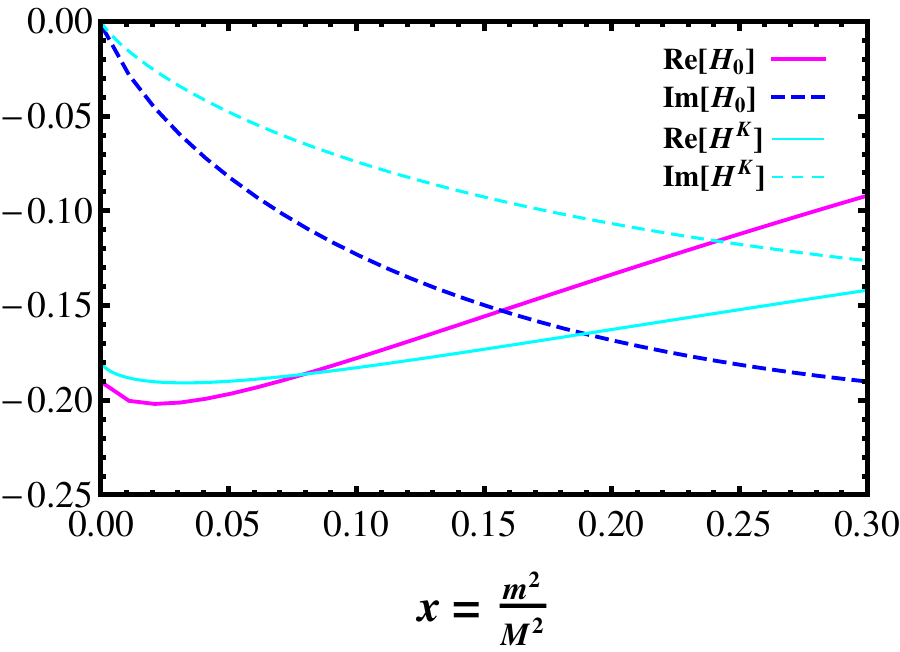}
\caption{\label{H0}The mass dependence of the dimensionless function $H_{0}$. The function $\hat{H}^{k}$ was obtained in Ref.~\cite{Korner:1982vg}.}
\end{figure}

In the limit of $x=m^{2}/M^{2} \rightarrow 0$,
\begin{align*}
g_{1}(\xi)&\rightarrow    0,\\
g_{2}(\xi)&\rightarrow   \frac{1}{1-\xi}\left(2 \ln^{2} x+ 4 i \pi \ln x  +\frac{ \pi ^2}{3}-8 i \pi  \ln \frac{2}{1+\xi}+4 \mathrm{Li}_{2}(-\xi)- 4\mathrm{Li}_{2}(\xi)\right),\\
g_{3}(\xi)&\rightarrow    0,\\
g_{4}(\xi)&\rightarrow  0,\\
g_{5}(\xi)&\rightarrow \frac{1}{1-\xi}\left(-2 \ln^{2} x-4 i \pi \ln x +\frac{2 \pi ^2}{3}+8 i \pi  \ln \frac{2}{1+\xi}\right)\\
g_{6}(\xi)&\rightarrow \frac{2 }{1-\xi}\bigg(-\ln 4 +(1-\xi) \ln(1-\xi)+(1+\xi) \ln(1+\xi)\bigg),\tag{5}
\end{align*}
and
\begin{align*}
     I_{0}(u)&\rightarrow \frac{8}{1-\xi^2}\left(\frac{\pi^2}{4}-\ln 2-\frac{\xi }{2} \ln\left(\frac{1-\xi}{1+\xi}\right)+\frac{1}{2} \ln\left(1-\xi^2\right)+\xi\mathrm{Li}_{2}(-\xi)-\xi \mathrm{Li}_{2}(\xi)\right),\tag{6}
\end{align*}
which reproduces the weight function $\mathcal{W}(u)$ in the Eq.~(11) of the work by K\"{u}hn~\cite{Kuhn:1983yr}.
Performing the convolution integral between the loop function $I_{0}(u)$ and the asymptotic DA,
we display the mass dependence of $H_{0}$ in Fig.~\ref{H0}. For comparison, we also present the dimensionless function $H^{K}$ obtained in Ref.~\cite{Korner:1982vg}.
As known, the contribution of on-shell gluons in the partial wave $0^{-+}$ is suppressed by a factor of $m^{2}/M^{2}$~\cite{Krammer:1978qp,Billoire:1978xt}, and our results for the absorptive part of the function $H_{0}$, which represents the contribution of on-shell gluons, indeed follow this character in the limit of $m^{2}/M^{2}\rightarrow0$. However,
the dispersive part of the function $H_{0}$, which represents the contribution of virtual gluons, is not suppressed in the limit of $m^{2}/M^{2}\rightarrow0$. While the matrix elements $\langle0|G_{\mu\nu}^a\tilde{G}^{a,\mu\nu}|\eta^{(\prime)}\rangle$ are twist-4 effects which are suppressed by the factor of $m^{2}/M^{2}$ relative to the leading twist terms. That is the reason
why the mass dependence of the decay widths obtained in this work differs from that obtained in Ref.~\cite{Novikov:1979uy} by an additional factor $M^{4}/m^{4}$.
In additions, when the factor $m^{2}/M^{2}$ is not very small, the suppression for the absorptive part is no longer operative. For example, the dimensionless function $H_{0}^{\eta}=H_{0}{\mid}_{m=m_{\eta}}$ is dominated by its dispersive part, while the dispersive part and the absorptive part of the $H_{0}^{\eta^{\prime}}=H_{0}{\mid}_{m=m_{\eta^{\prime}}}$ are comparable.



\end{document}